\documentclass[acus]{JAC2000}

\usepackage{graphicx}

\begin{document}
\title{
\flushright{T17}
\\[15pt]
\centering
Preliminary trigger studies for the PEP-N detector}

\author{E. Pasqualucci, INFN, sez. di Roma, P.le A. Moro 2,
        I-00185, Italy}
 
\maketitle

\begin{abstract}

A very preliminary study for a trigger for the PEP-N
experiment is presented. Its aim is to show the feasibility
of a very efficient trigger for multihadronic events
based on energy releases in the
electromagnetic calorimeter.
Though the efficiency of such a trigger is very high
and simple topology requirements
can be applied to reduce cosmic background, an
accurate study of machine background is going to be
made to provide final trigger design and performance.

\end{abstract}

\section{Introduction}

The PEP-N experiment needs a very efficient trigger
in order to minimize the systematics.
In particular, the exclusive approach\cite{Bett} is very demanding
from this point of view, requiring a well known efficiency
for each channel.
Moreover, luminosity measurement\cite{Mandel}
requires full efficiency for bhabha events.

On the other hand, background rejection is a crucial 
point\cite{Kell}.
At this stage, machine background has not been extensively
studied.
In this paper, a very preliminary evaluation of trigger
efficiency for multihadronic channels is presented.
Proposed trigger criteria do not still take into
account the effect of machine background; aim of this study
is to demonstrate the possibility to build a very efficient
trigger and to apply some simple topological criterium to 
cut cosmic background without affecting
the efficiency.
Final trigger design will be heavily driven by the
machine background characteristics.

In the following, some results are shown for a simple
trigger based on energy deposits in the electromagnetic calorimeter.

\section{Multihadronic events}

The trigger efficiency for multihadronic events has been studied
generating some thousands of 
events for each channel and following the particles
up to the electromagnetic calorimeter.
The energy release have been simulated according to the
experimental data from the KLOE calorimeter\cite{Kcal} for forward
and barrel modules. An accurate montecarlo describing
calorimeter modules, including lead, fibers and glue,
has been used to simulate the thin pole and backword modules,
comparing results with the KLOE ones.
Impact angles and border effects have been taken into account.
Signals coming from low energy incident particles have been weighted
according to the proper efficiency.

As a minimum bias trigger criterium, two energy deposits over a given
threshold have been required. A few photo-electrons can be seen by our
calorimeter, that is fully efficient yet for 40 MeV
incident gammas. In our case, we accepted events with
at least two releases over 2 MeV in fibers (corresponding to
about 15-20 MeV incident gammas). 
This requirement is very realistic, being based on the behaviour
of the 4m long modules of the KLOE calorimeter,
which is now working in Frascati.
The different energy deposit in the
pole and backword calorimeter, which has a different
thickness, has been taken into account looking at the results
of the simulation.
Low energy signals have been weighted taking into account
the efficiency of the calorimeter for the appropriate particle at that
energy.

\begin{figure}[htb]
\centering
\includegraphics[width=75mm]{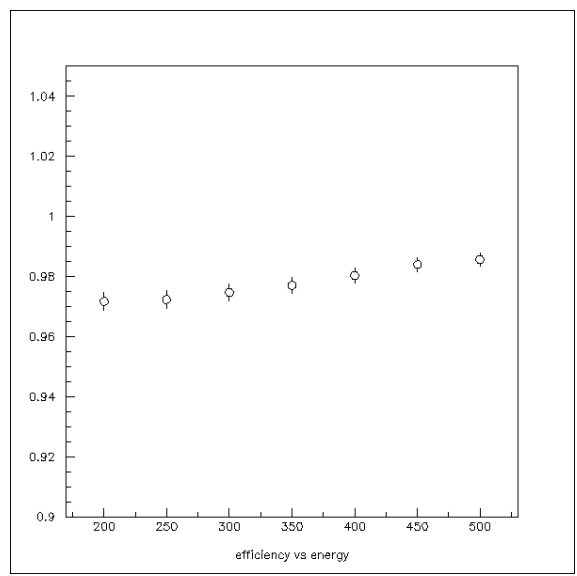}
\caption{Trigger efficiency for the reaction
         $e^+e^- \rightarrow k^+k^- \pi^+ \pi^-$.}
\label{kkpp}
\end{figure}

In figure \ref{kkpp} the results for the reaction
$e^+e^- \rightarrow k^+k^- \pi^+ \pi^-$ are shown as
an example.
The trigger efficency for this channel
is plotted as a function of the VLER\cite{mach} beam energy (in MeV).

The trigger efficiency for the most significant hadronic
channels at two energies is shown in table \ref{mheff}.

\begin{table}[htb]
\centering
\begin{tabular}{|l|c|c|}
\hline
\textbf{Reaction} & \textbf{VLER energy (MeV)} & \textbf{Efficiency} \\
\hline
$\pi^+\pi^-2\pi^0$     & 200 & .999 \\
                       & 350 & 1 \\
\hline
$2\pi^+2\pi^-$         & 200 & .991 \\
                       & 350 & .995 \\
\hline
$\pi^+\pi^-4\pi^0$     & 200 & 1 \\
                       & 350 & 1 \\
\hline
$2\pi^+2\pi^-2\pi^0$   & 200 & 1 \\
                       & 350 & 1 \\
\hline
$k^+k^-\pi^+\pi^-$     & 200 & .972 \\
                       & 350 & .977 \\
\hline
\textbf{global}        & \textbf{200} & \textbf{.994} \\
                       & \textbf{350} & \textbf{.995} \\
\hline
\end{tabular}
\caption{Minimum bias trigger efficiency for the most significant
hadronic channels at 1.58 and 2.1 Gev in the centre of mass.}
\label{mheff}
\end{table}

\section{Bhabha events}

Trigger efficiency on bhabha channel is very important,
in particular on the events concurring to the luminosity measurement.

\begin{figure}[htb]
\centering
\includegraphics[width=75mm]{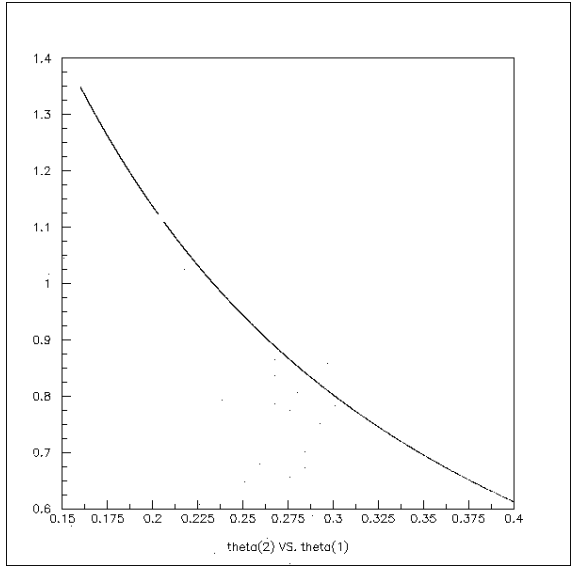}
\caption{Angular correlation of bhabha events.}
\label{bcorr}
\end{figure}

The production angles for electron and positron in a bhabha event
in the laboratory system
are strongly correlated (figure \ref{bcorr}).

If we analise the events with a particle
between $\theta=0.3$ and $\theta=0.4$
in the laboratory system (the ones used
for the luminosity measurement,
according to M. Mandelkern's document \cite{Mandel}),
they are, depending on the energy of the electron beam, or both in
the forward calorimeter, or one in the forward
and the other in the barrel or pole.

\begin{figure}[htb]
\centering
\includegraphics[width=75mm]{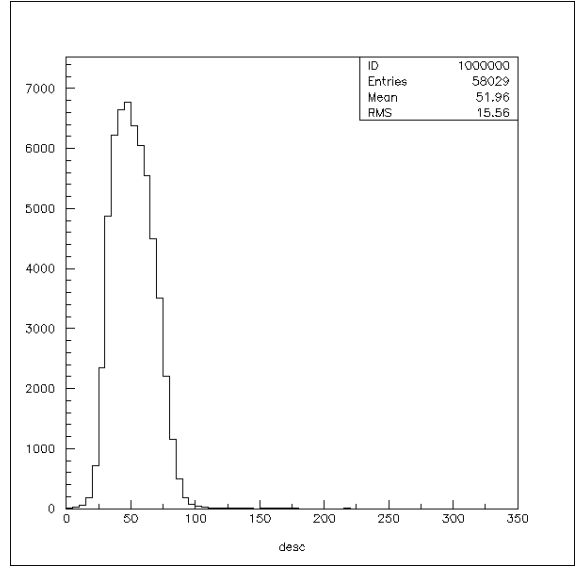}
\caption{Energy deposited in fibres by the
         low energy particle in a bhabha event.}
\label{bdepo}
\end{figure}

In the worst case, the lowest energy particle
will hit the thin pole calorimeter.
The distribution of released energy for that particles is shown in figure
\ref{bdepo}.

With our threshold, we have full efficiency.
Also with higher thresholds
($\sim$ 10 MeV deposited per particle) the efficiency
is almost full.

\section{Cosmic background}

Though they are not the most important source of background,
it is useful to show that cosmic rays can be reduced
just applying simple topological criteria.
Due to the boost in the laboratory system and the
angular coverage of the detector, a very few events
produce only two hits in the barrel or pole calorimeter.
If we choose not to trigger on these events,
a reduction factor at least 3 in cosmic ray background is expected.
 
\begin{figure}[htb]
\centering
\includegraphics[width=75mm]{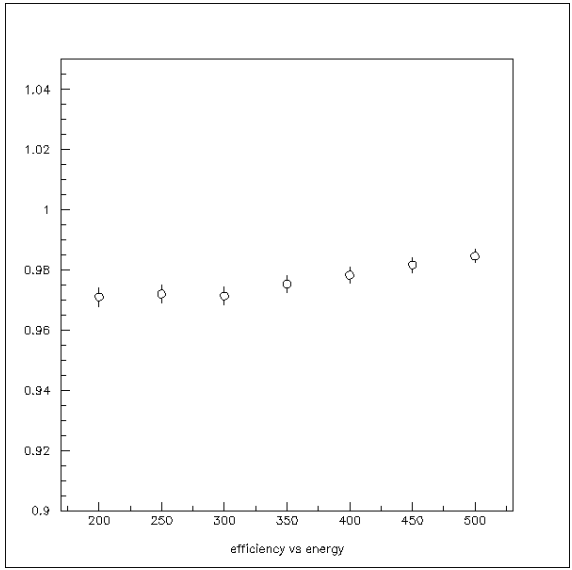}
\caption{Trigger efficiency for the reaction
         $e^+e^- \rightarrow k^+k^- \pi^+ \pi^-$ with
         cosmic background reduction.}
\label{coskkpp}
\end{figure}

According to the plot of figure \ref{bcorr},
luminosity bhabha cannot be affected by this requirement.
In figure \ref{coskkpp} the trigger efficiency for the reaction
$e^+e^- \rightarrow k^+k^- \pi^+ \pi^-$ after
cosmic background reduction is shown. 

A comparison with
figure \ref{kkpp} shows that the effect of the topological criterium
is negligible. This consideration applies for all the
hadronic channel.
The overall efficiency for hadronic processes is again
.994 at 200 MeV of VLER energy and .995 at 350 MeV,
to be compared with the results in table \ref{mheff}.

\section{Final considerations and further work}

The overall trigger rate at a luminosity
$10^{31}$ cm$^{-2}$ s$^{-1}$ can be anticipated to be the sum
of the contributions in table \ref{frq}.

Provided we can take under control the
trigger rate due to machine background,
these rates are not challenging for a modern
data acquisition system, considering an expected
event size of some kByte.
Data can be processed and filtered online both
with an hardware or a software second level trigger,
thus lowering the background level.

\begin{table}[htb]
\centering
\begin{tabular}{|l|c|}
\hline
hadrons + muons                     & $\sim$ 1 Hz \\
\hline
bhabha (cutting on $\theta$ = 0.16) & $\sim$ 20 Hz \\
\hline
cosmics                             & $\sim$ 150-200 Hz \\
\hline
machine background                  & ? \\
\hline
\end{tabular}
\caption{trigger rates.}
\label{frq}
\end{table}

Work is in progress to study the effect of the machine background.
In general, we can expect many background events 
firing the trigger.
Background rejection could be improved adding some topological
information or requiring more hits in the calorimeter for
many-body events. 
In the latter case, angular correlation provides the 
way to accept two-body events.

An accurate study of machine background characteristics
has been started to provide final trigger design and performance.

\end{document}